\documentclass[11pt,a4paper]{article}
\usepackage{jheppub}
\usepackage{graphicx}
\usepackage{epsfig}
\usepackage{amssymb}
\usepackage{amsmath}
\usepackage{float}
\usepackage{subfigure, amsmath,graphicx}

\newcommand{\bea}{\begin{eqnarray}}
\newcommand{\eea}{\end{eqnarray}}
\newcommand{\bean}{\begin{eqnarray*}}
\newcommand{\eean}{\end{eqnarray*}}
\newcommand{\nn}{\nonumber \\}

\def\W #1{\widetilde{#1}}
\def\WH #1{\widehat{#1}}

\def\eref#1{(\ref{#1})}

\def\a{{\alpha}}

\def\b{{\beta}}

\def\Spba #1{ \left[ #1 \right\rangle}
\def\Spbb #1{ \left[ #1 \right]}
\def\Spaa #1{ \left\langle #1 \right\rangle}
\def\Spab #1{ \left\langle #1 \right]}

\def\ket#1{\left| #1\right\rangle}

\def\bket#1{\left| #1\right]}

\def\Label#1{\label{#1}%
  \smash{\hbox to0pt{\raise1ex\hbox{\tiny[#1]}\hss}}}

\title{Graphs, determinants and gravity amplitudes}

\author[a,b,d]{ Bo Feng}

\author[c]{Song He}

\affiliation[a]{Zhejiang Institute of Modern Physics, Zhejiang
University, Hangzhou, 310027, P. R. China}

\affiliation[b]{Center of Mathematical Sciences, Zhejiang
University, Hangzhou, 310027, P. R.  China}

\affiliation[c]{Max-Planck-Institut f\"ur Gravitationsphysik, Am
M\"uhlenberg 1, 14476 Potsdam, Germany}

\affiliation[d]{Niels Bohr International Academy and Discovery
Center, The Niels Bohr Institute, Blegdamsvej 17, DK-2100,
Copenhagen, Denmark}

\emailAdd{fengbo@zju.edu.cn}

\emailAdd{songhe@aei.mpg.de}

\date{\today}
\abstract{We apply the matrix-tree theorem to establish a link between various diagrammatic and determinant expressions, which naturally appear
in scattering amplitudes of gravity theories. Using this link we are able to give a general graph-theoretical formulation for the tree-level
maximally-helicity-violated gravity amplitudes. Furthermore, we use the link to prove two identities for half-soft functions of gravity
amplitudes. Finally we recast the diagrammatic formulation of one-loop rational part of $\mathcal{N}=4$ supergravity into a matrix form. }

\begin{document}
\maketitle

\section{Introduction}

In recent years we have witnessed tremendous progresses in understanding scattering amplitudes in gauge and gravity theories~\cite{ampreview}.
Generally, Britto-Cachazo-Feng-Witten (BCFW)~\cite{Britto:2004ap,Britto:2005fq} recursion relations can be used to obtain tree-level amplitudes,
while loop amplitudes can be determined by generalized-unitarity method~\cite{Bern:1994zx,Bern:1994cg}. In particular, amplitudes in
$\mathcal{N}=4$ super--Yang-Mills (SYM), which is probably our best studied example, possess even more beautiful structures and simplicities,
and powerful tools such as the Grassmannian formulation~\cite{Grassmannian} have been developed recently to unravel them. Moreover, the theory
is believed to be integrable in the planar limit~\cite{intreview}, which has made it possible to determine all-loop scattering amplitudes in its
planar sector~\cite{all-loop}.

Via the celebrated Kawai-Lewellen-Tye (KLT) relations~\cite{Kawai:1985xq}, tree-level amplitudes in gravity can be constructed by recycling the
corresponding gauge theory amplitudes. More recently, Bern, Carrasco and Johannsson (BCJ) proposed a surprising duality between the color and
kinematics of color-dressed amplitudes in gauge theories~\cite{Bern:2008qj}, which can be ``squared" to give gravity
amplitudes~\cite{Bern:2010yg}~\footnote{Although these relations can be derived in string theory~\cite{BCJstring}, they can also be proved
within field theories~\cite{Bern:2010yg,BCJfieldtheory,KLTfieldtheory}. Similar relations have been discovered in three-dimensional theories
where string interpretations are unknown~\cite{Bargheer:2012gv}.}. They also conjectured that similar construction holds for gravity amplitudes
at loop level~\cite{Bern:2010ue}, which has played a key role in the recent heroic calculations of multi-loop amplitudes in $\mathcal{N}=8$
supergravity (SUGRA)~\cite{SUGRAUV}. Despite these remarkable progresses, by far most formulations, which use gauge theory amplitudes, are
rather complicated because the number of terms involved grows as factorial, and they often obscured some nice properties of gravity amplitudes
(e.g. in these formulations SUGRA amplitudes are definitely not ``simpler'' than the SYM ones~\cite{ArkaniHamed:2008gz}). With this in mind, in
the present note we intend to further explore the structure and simplicity of gravity amplitudes, without any reference to gauge theory.

For the purpose of revealing the structure, it is already enough to look at the simplest example: the maximally-helicity-violated (MHV) gravity
amplitudes, and a few interesting steps have been taken in this direction. At tree level, although ``old-fashioned'' expressions for MHV gravity
amplitudes have been known since~\cite{Berends:1988zp}, there are two recent formulations for which have no resemblance to gauge theory
amplitudes: the formula by Nguyen, Spradlin, Volovich and Wen (NSVW)~\cite{Nguyen:2009jk}, and that by Hodges~\cite{Hodges:2012ym}. The former,
originally derived from the link representation, writes the amplitude as a sum over labeled tree graphs,  while the latter expresses it as a
single determinant~\footnote{We would like to thank B.J. Spence about the observation of relations between these two formula.}~\footnote{While
the note is being written, two related, but different formula inspired by twsitor-string have been proposed to give all N$^k$MHV tree amplitudes
~\cite{Cachazo:2012da}~\cite{Cachazo:2012kg}, both used Hodges' determinant as a prototype.}. One-loop amplitudes of $\mathcal{N}=8,6,4$ SUGRA
have also been extensively studied, and a nice approach is to construct them using soft and collinear
factorizations~\cite{Dunbar:2010fy,Dunbar:2011dw,Dunbar:2012aj}. In particular, the rational part of one-loop MHV amplitudes in $\mathcal{N}=4$
SUGRA, which can not be determined from unitarity method, has been computed and expressed as ``one-loop" labeled graphs similar to the NSVW
formula for tree amplitudes~\cite{Dunbar:2012aj}, based on an identity of half-soft functions, which we will give a simple proof.

In this note, we consider a possibly universal theme underlying these formulations, namely gravity amplitudes can be naturally written in two
equivalent representations: \textsl{graphs and determinants}. In section~\ref{graph-det} we recall the matrix-tree theorem, which relates
determinants constructed from a labeled graph to its spanning trees/forests. The theorem immediately yields a graph-theoretical interpretation
of Hodges' formula, which includes but also generalizes the NSVW formula, as shown in section~\ref{tree}. The interesting connection between
graphs, determinants and gravity amplitudes goes beyond the example. In section~\ref{half-soft}, we find that the half-soft and soft-lifting
functions, which play an important role in constructing gravity amplitudes, can be defined naturally in terms of graphs/determinants, and some
identities can be nicely proved using these formulations. As another example, in section~\ref{one-loop}, we apply the theorem to the one-loop
rational part of $\mathcal{N}=4$ SUGRA MHV amplitudes, and rewrite the diagrammatic formula of~\cite{Dunbar:2012aj} into a determinant form.

\section{Graphs and Determinants}~\label{graph-det}

Let us recall a few definitions in graph theory. A (simple) graph $G=(V,E)$ comprises a set $V$ of vertices, and a set $E$ of edges, where each
edge $e$ is a pair\footnote{The graph is called directed if the pairs are ordered, and undirected if they are unordered. In the following we
consider undirected graphs.} of two different vertices $v, w$, $e= v w$ (we say $v$ and $w$ are adjacent, $v\sim w$), and there are no more than
one edge between two vertices. A tree is a connected graph without cycles, and a forest is a disjoint union of trees.

\begin{figure}\centering
\includegraphics[height=4cm]{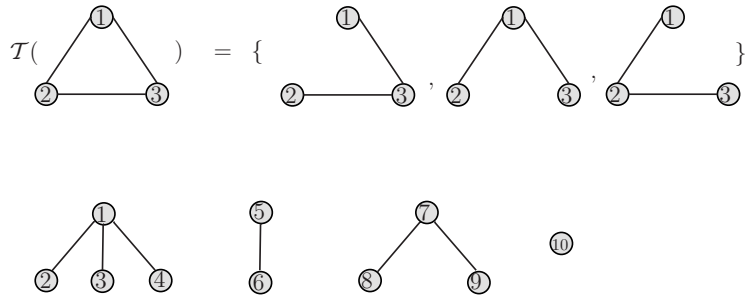}
\caption{There are 3 labeled spanning trees of the complete graph with 3 vertices, $K_3$ (top). An example of labeled forests with 10 vertices,
which has 4 trees; if one chooses e.g. vertices 1,5,7,10 to be the roots, it becomes a rooted forest (bottom).} \label{fig:tree_forest}
\end{figure}

Furthermore, a tree is called rooted if one vertex is designated as a root, and a rooted forest is a forest of rooted trees. A spanning tree of
a connected graph $G$ is a tree with all the vertices and a subset of the edges of $G$, and similarly for spanning forests. See
Fig.~\ref{fig:tree_forest} for examples.

\subsection{Spanning trees}

There is an interesting theorem connecting the determinant of certain matrices associated with a graph, and its spanning
trees~\cite{Stanleybook}.

For a connected, simple graph $G$ with vertices $V=\{v_1,...v_n\}$, one can assign a weight $\psi_{i j}$ to each edge $e=v_i v_j$, and define a
$n\times n$ {\bf weighted Laplacian matrix} $W(G)$,
\bea [W(G)]_{ij}=\left\{  \begin{array}{ll}  \sum_{v_k\sim v_i} \psi_{ik} & \textrm{if}~i=j
\\  -\psi_{ij} & \textrm{if}~i\sim j\\ 0 & \textrm{otherwise}\end{array}\right.~~~~~\label{W-Lap-G}\eea
Note that the sum of elements in a row or a column vanishes, so the
matrix is degenerate. However, we could pick a vertex $v_i$ and
consider the minor corresponding to the element $W[G]_{i}^{i}$,
where we use the lower index to denote the $i$-th row having been
deleted and upper index $i$ to denote the $i$-th column having been
deleted.

{\bf Matrix-Tree Theorem I}: The determinant of the submatrix obtained by deleting the $i$-th row and $i$-th column is independent of $i$, and
it is given by,
 \bea \boxed{|W(G)|_{i}^i= \sum_{T\in {\cal T}(G)} \left(
\prod_{e=(v_i v_j)\in E(T)} \psi_{ij}\right)}, ~~~\label{Weighted-WG-det}\eea
where the sum is over all spanning trees $T\in {\cal T}(G)$ and the product is over all edges of $e\in T$.

For a labeled complete graph $K_n$, where any two vertices are connected by exactly one edge, spanning trees are all possible trees with $n$
vertices, and it is straightforward to use the theorem to enumerate them. Setting $\psi_{i j}=1$, thus $[W(G)]_{ii}=n{-}1$, and we recover
Cayley's formula which gives the number of all trees with $n$ vertices, $|{\cal T}(K_n)|=n^{n-2}\label{Cayler-formula}$. We will see in the next
section that, with a suitable choice of $\psi_{i j}$, the theorem reduces Hodges' formula to the tree-diagram formula in ~\cite{Nguyen:2009jk}.

\subsection{Forests}

The above theorem can be generalized to the case of rooted forests, which turns out to be the general diagrammatic expansion of Hodges' formula.
Given any simple graph $G$ with $n$ vertices, we consider the spanning forests of $G$ and define the same weighted Laplacian matrix $W[G]$. Now
if we pick a subset of vertices $I$, there is a a theorem relating the corresponding minor and forests with roots in $I$.

{\bf Matrix-Tree Theorem II}: Let us denote the set of rooted forests with $r$ trees, which have roots with labels $I=\{i_1,...,i_r\}$, by
${\cal F}_I(G)$, then the determinant of the submatrix after deleting the rows and columns with indices $i\in I$ is given by,
\bea \boxed{|W(G)|^{i_1...i_r}_{i_1...i_r}=\sum_{F\in {\cal F}_I(G)} \left(\prod_{e=v_i v_j\in E(F)} \psi_{i
j}\right).}~~~~\label{forest-gen}\eea

We illustrate the proof of both theorems by induction for complete graph $K_n$, and the proof for other graphs follows similarly. For $K_n$ (the
case for $K_2$ is trivial), let us focus on the case $r=1$ (Theorem I), and without loss of generality we choose $I=\{n\}$, ${\cal
F}_{\{n\}}(K_n)={\cal T}(K_n)$. In this case, $\mathcal{T}(K_n)$ can be obtained by consider all possible ways of connecting vertex $v_n$ to
vertices in spanning forests of $K_{n-1}$.

The simplest case is when $n$ is connected to one vertex $p_1$, where the set of spanning forests of $K_{n-1}$ with one root $p_1$ is exactly
${\cal F}_{\{p_1\}}(K_{n-1})={\cal T}(K_{n-1})$, then we have a contribution for the RHS, \bea \sum^{n{-}1}_{p_1=1} \psi_{n p_1}\sum_{F\in {\cal
T}(K_{n-1})}\left(\prod_{e=v_iv_j\in E(F)} \psi_{i j}\right)=\sum^{n-1}_{p_1=1} \psi_{n p_1} |W(K_{n-1}|^{p_1}_{p_1}, \eea where we have used
the induction assumption for $n-1$ vertices (with $r=1$). Generally when $n$ is connected to $r$ vertices, $p_1,p_2,...,p_r$, in order to obtain
a tree of $n$ vertices, we need to consider the forests of $K_{n-1}$ with $r$ trees, which have exactly roots $p_1,p_2,...,p_r$, and the RHS
reads,\bea \sum_{F\in {\cal F}_I(G)} \left(\prod_{e=v_iv_j\in F} \psi_{i j}\right)=\sum^{n{-}1}_{r=1}\sum_{p_1<...<p_r}\prod^r_{k=1} \psi_{n
p_k} |W(K_{n-1})|^{p_1...p_r}_{p_1...p_r}.\eea On the other hand, if we consider the submatrix after deleting $n$-th row and column,
$W(K_n)^n_n$, the determinant can be expanded in $\psi_{n i}$, which only appear in the diagonal. The zeroth order is $|W(K_{n-1})|=0$, and the
$r$-th order is the same as above, \bea |W(K_n)|^n_n=\sum^{n{-}1}_{r=1}\sum_{p_1<...<p_r}\prod^r_{k=1} \psi_{n p_k}
|W(K_{n-1})|^{p_1...p_r}_{p_1...p_r}~~\label{det-expansion}.\eea

Therefore we have proved the $r=1$ case for $n$ vertices. Essentially the same proof holds for $r=2,...,n$, by connecting each of the $r$ roots
to forests with $n{-}1$ vertices, and that completes our inductive proof. We remark that~\ref{det-expansion} was also used by Hodges to prove
his formula using modified BCFW recursion relations given in~\cite{Hodges:2011wm}, and we have seen that each term now has a nice
graph-theoretical/combinatoric interpretation.
\section{The tree-level gravity MHV amplitude}~\label{tree}

Our first application of the theorem is to understand relations between two formula of tree-level gravity MHV amplitude: the Hodges'
determinant formula~\cite{Hodges:2012ym} and the NSVW tree-diagram formula~\cite{Nguyen:2009jk}. 

\subsection{Hodges' determinant formula}

We will present two matrix forms: the original form by Hodges, and a more symmetric form related to it by simple transformations. Hodges'
original form is {\sl manifestly independent of the choice of auxiliary spinors $x,y$}, while the new one is {\sl manifestly related to the
graphic representation}.

Using $a_k\equiv \Spaa{k|x}\Spaa{k|y}$ where $x,y$ are two auxiliary
spinors, the element of the matrix defined by Hodges, $\Phi$, is
given by\footnote{It is worth to emphasize that the sign here is
different from the one given by Hodges. We find it more convenient
since it is consistent with the matrix-tree thereom.}
\bea \phi_{i}^j= -{\Spbb{i|j}\over \Spaa{i|j}}~~for~i\neq j,~~~~\phi_i^i =\sum_{j\neq i} \phi_i^j {a_j\over a_i}~,~~~\label{Hodges-element} \eea
where lower (upper) indices are for rows (columns).  In the matrix
form, \eref{Hodges-element} reads,
\bea \Phi=\left( \begin{array}{ccccc} \sum_{j\neq 1} \phi_1^j
{a_j\over a_1} &  -\phi_1^2  & \ldots   & -\phi_1^{n-1} & -\phi_1^n \\
-\phi_2^1 & \sum_{j\neq 2} \phi_2^j {a_j\over a_2}  & \ldots   & -\phi_2^{n-1} & -\phi_2^n  \\
\vdots &\vdots   & \ddots  & \vdots &\vdots \\
-\phi_{n-1}^1 & -\phi_{n-1}^2  & \ldots  & \sum_{j\neq n-1} \phi_{n-1}^j {a_j\over a_{n-1}}
 & -\phi_{n-1}^{n}\\
-\phi_{n}^1 & -\phi_{n}^2 &  \ldots  & -\phi_n^{n-1} & \sum_{j\neq n} \phi_{n-1}^j {a_j\over a_{n}}
\end{array} \right)~.~~\label{Phi-H}\eea
Although in the definition of diagonal elements auxiliary spinors, $x,y$, have been introduced, momentum conservation implies that the matrix is
independent of the choice. In addition, one can check with momentum conservation that
\bea \Phi\cdot ( \Spaa{1|\a}\Spaa{1|\b},\Spaa{2|\a}\Spaa{2|\b},...,\Spaa{n|\a}\Spaa{n|\b})=0 \eea
for arbitrary spinors $\a,\b$. Since any spinor can be expanded using a basis of two spinors, the space of the above null vectors has dimension
three.

Since $\Phi$ has rank $n{-}3$, only minors with dimension $d\leq n{-}3$ do not vanish. Let us use $(\Phi)_{ijk}^{rst}$ to denote the matrix
obtained by removing rows $i,j,k$ and columns $r,s,t$ from $\Phi$ and $|\Phi|_{ijk}^{rst}$ for its determinant (we require $i<j<k$, $r<s<t$ to
avoid ambiguities),the MHV tree amplitude of gravity is given by \cite{Hodges:2012ym}
\bea {\cal M}_n & = &  (-)^{i+j+k+r+s+t} c^{ijk} c_{rst}|\Phi|_{ijk}^{rst}~,~~~\label{MHV-H} \eea
where we have suppressed the momentum-conservation delta-functions etc., and \bea c_{ijk}=c^{ijk}= {1\over \Spaa{i|j}\Spaa{j|k}\Spaa{k|i}}~.\eea

It has been proved in \cite{Hodges:2012ym} that \eref{MHV-H} is totally symmetric under permutation of $n$ particles. This is related to the
fact that $\Phi$ has rank $n{-}3$, so  different $(n{-}3)\times (n{-}3)$ minor has following relation
\bea (-)^{i+j+k+r+s+t} c^{ijk} c_{rst}|\Phi|_{ijk}^{rst}= (-)^{\W i+\W j+\W k+\W r+\W s+\W t} c^{\W i\W j\W k} c_{\W r \W s\W t}|\Phi|_{\W i\W
j\W k}^{\W r \W s\W t}~.~~~\label{Phi-minor-rel}\eea
%

\subsection{NSVW formula}

Having reviewed Hodges' formula, we want to demonstrate its relation to NSVW formula. To do so, matrix form \eref{Phi-H} is not so convenient
and we need to define following new matrix form
\bea \Psi= A \cdot \Phi \cdot A~,~~~~\label{Psi-H}\eea
where matrix $A$ has only diagonal elements and it given by $A={\rm diag} (a_1, a_2,...,a_n)\label{diag-A}$. Writing up into matrix form we have
\bea \Psi=\left( \begin{array}{ccccc} \sum_{j\neq 1} \phi_1^j {a_j} a_1 &  -\phi_1^2 a_1 a_2 &
\ldots & -\phi_1^{n-1}a_1 a_{n-1} & -\phi_1^n a_1 a_n\\
-\phi_2^1 a_1  a_2& \sum_{j\neq 2} \phi_2^j {a_j} a_2&
\ldots & -\phi_2^{n-1}a_2 a_{n-1} & -\phi_2^n a_2 a_n\\
 \vdots &\vdots  & \ddots & \vdots &\vdots \\
-\phi_{n-1}^1  a_1 a_{n-1}& -\phi_{n-1}^2 a_{n-1} a_2  & \ldots & \sum_{j\neq n-1} \phi_{n-1}^j {a_j} a_{n-1}
 & -\phi_{n-1}^{n} a_{n-1} a_n\\
-\phi_{n}^1  a_1  a_{n}& -\phi_{n}^2  a_2 a_{n}& \ldots & - \phi_n^{n-1} a_{n-1}a_{n}& \sum_{j\neq n} \phi_{n-1}^j {a_j} a_n
\end{array} \right)~~~\label{New-Phi}\eea
Matrix $\Psi$ has a nice property: the sum of each row (or each column) is zero. This is the character of Laplacian matrix used in the
matrix-tree theorem.

Using the relation \eref{Psi-H}, it is easy to see that
\bea |\Psi|_{ijk}^{rst}= |A|_{ijk}^{ijk} |\Phi|_{ijk}^{rst} |A|_{rst}^{rst}= \left({a_i a_j a_k a_r a_s a_t\over (\prod_{i=1}^n
a_i)^2}\right)^{-1} | \Phi|_{ijk}^{rst}~,~~~\label{Phi-Psi-rel}\eea
thus the gravity MHV amplitude can be written as
\bea {\cal M}_n & = &  (-)^{i+j+k+r+s+t} c^{ijk} c_{rst}{a_i a_j a_k a_r a_s a_t\over (\prod_{i=1}^n a_i)^2} |
\Psi|_{ijk}^{rst}~.~~~\label{MHV-H-Psi} \eea

\begin{figure}\centering
\includegraphics[height=2.5cm]{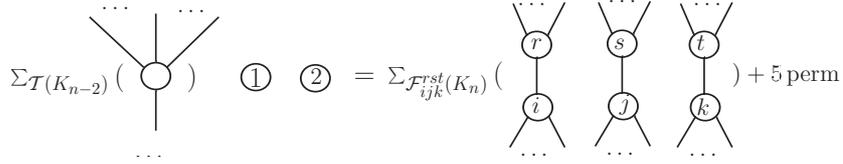}
\caption{Diagrammatic expansions of Hodges' determinant formula for gravity MHV tree amplitudes. A special choice (reference points $1,2$) gives
the NSVW formula as the sum of weighted spanning tree (left). The most general diagrammatic expansion of the formula is the sum of weighted
forests with 3 trees, which contain $\{i,r\},\{j,s\},\{k,t\}$ (or $S_3$ permutations) respectively (right).} \label{fig:gravity_tree}
\end{figure}

Now we can see relation to NSVW formula clearly. We can take $s=j=x=n{-}1$, $t=k=y=n$, then the last two rows and two columns of $\Psi$ vanish,
and matrix $\Psi$ is reduced to a $(n{-}2)\times (n{-}2)$ matrix effectively. Thus \eref{MHV-H-Psi} becomes
\bea {\cal M}_{n} = {  (-)^{i+r} \over \Spaa{n-1|n}^2 \prod_{i=1}^{n-2} a_i^2} | \WH \Psi|_{i}^{r}~,~~\label{NSVW-det} \eea
where we used the $\WH \Psi$ to denote the reduced $(n{-}2)\times (n{-}2)$ matrix. For this reduced matrix we have $(-)^{i+r}| \Psi|_{i}^{r}=
(-)^{\W i+\W r}| \Psi|_{\W i}^{\W r}\label{Psi-minor-rel}$, thus
\bea {\cal M}_{n} = {  1 \over \Spaa{n-1|n}^2 \prod_{i=1}^{n-2} a_i^2} | \WH \Psi|_{r}^{r}~.~~\label{NSVW-det-1}\eea
The expansion of \eref{NSVW-det-1} is exactly the NSVW tree-diagram formula by the matrix-tree theorem. See the left part of
Fig.~\ref{fig:gravity_tree}.

\subsection{The general diagrammatic expansion of Hodges' formula}

Now it is clear that by the matrix tree theorem I, NSVW tree-diagram expression is just a special case of Hodges' determinant formula after we
make the choice $s=j=x=n{-}1$, $t=k=y=n$. Here we would like to show the most general diagrammatic expansion of Hodges' formula, using the
theorem II and a bit generalizations.

For determinant $|\Psi|_{ijk}^{rst}$ with $i=r, j=s, k=t$, the diagrams are all rooted forest with three disconnected trees, whose roots are at
$i,j,k$. The weight of each edge is nothing but ${\Spbb{r|s}\over \Spaa{r|s}} (\Spaa{r|x}\Spaa{r|y})(\Spaa{s|x}\Spaa{s|y})$.

For the case $|\W \Phi|^{r j k}_{i j k}$ with $r\neq i$, we will need forests with three trees: the first tree must have nodes $r,i$, the
second, node $j$, and the third, node $k$. The sign of all terms are same.

For the case $|\W \Phi|^{r s k}_{i j k}$ where only one pair of indices coincide, all contributions will be divided into two kinds of graphes
with opposite sign. The first kind contains forests where $(i,r)$ are at the first tree, $(j,s)$ at the second, and $k$ at the third, and with
$r$ and $s$ interchanged for the second kind.

For the case  $|\W \Phi|^{r s t}_{i j k}$ where none of the indices coincide, all contributions will be divided into six kinds of graphes. If we
use $\sigma\in S_3$ to denote the permutation of indices $(r,s,t)$. Then each kind of graphes are given by forests with three trees, where
$(i,\sigma(r))$ at the first, $(j,\sigma(s))$ at the second, and $(k,\sigma(t)$ at the third, with a sign ${\rm sign}(\sigma)$. This is shown in
the right part of Fig.~\ref{fig:gravity_tree}.

\section{The half-soft function and soft-lifting function}~\label{half-soft}

Our second application of the graph-determinant connection is for the half-soft function and soft-lifting function, which are building blocks
for gravity amplitudes. For each function, we will give two equivalent definitions, using which some identities of half-soft function can be
proved in fairly straightforward way.

\subsection{The half-soft function}

The half-soft function was first defined in~\cite{Bern:1998sv}, which is used to give tree level MHV gravity amplitude. Recently, it is
understood that although looks different, half-soft function is equivalent to the MHV tree formula present in~\cite{Nguyen:2009jk}. Thus we can
immediately give two definitions of the half-soft function . The first is a diagrammatic expression,
\bea h(x, \{ 1,2,...,m\}, y) & = & {1\over \prod_{j=1}^m a_j^2} \sum_{ \textrm{trees}} \prod_{\textrm{edges}~(rs)} {\Spbb{r|s}\over \Spaa{r|s}}
a_r a_s,~~~~a_i=\Spaa{i|x}\Spaa{i|y},~~~\label{half-soft-graph}\eea
where the summation is over all spanning trees constructed by nodes $\{1,2,...,m\}$ and $x,y$ are auxiliary reference spinors. It is worth to
mention that in this definition, momentum conservation is not required, i.e., $\sum_{i=1}^m k_i\neq 0$ in general. The second one is to use
minors of the following matrix
\bea \Psi_i^j= -{\Spbb{i|j}\over \Spaa{i|j}} a_i a_j=-\phi_i^j a_i
a_j ~(i\neq j),~~~~\Psi_i^i=\sum_{j=1,j\neq i}^m
\Psi_i^j~,~~~\label{matrix-psi}\eea
and $h$ is written as
\bea h(x, \{ 1,2,...,m\}, y)= {1\over \prod_{j=1}^m a_j^2}|\Psi|_r^r= {1\over \prod_{j=1}^m a_j^2}||\Psi||,~~~\label{half-soft-matrix} \eea
where we have used the notation $||\Psi||\equiv |\Psi|_r^r$ to emphasize the independence of $r$.

These two definitions \eref{half-soft-graph} and \eref{half-soft-matrix} can be used to understand properties of half-soft function. The first
one is the soft limit behavior,
\bea h(x, M,y)|_{k_m\to 0} \to -{\cal S}_m (x, M, y) h(x, M-\{ m\}, y)~,~~~\label{h-soft-limit}\eea
where the {\sl half-soft factor} is defined as
\bea {\cal S}_m (x, M, y) ={-1\over a_m }\sum_{j\in M-\{m\}} {\Spbb{j|m}\over \Spaa{j|m}} a_j~.~~~\label{half-soft-factor}\eea
To see it, noticing that in \eref{half-soft-matrix} we can take minor by deleting any row and column, so to make things simpler, we take the
minor by deleting the $m$-th row and $m$-th column. In the remaining submatrix, only diagonal element has dependence on particle $m$ through
combination $\left({\Spbb{j|m}\over \Spaa{j|m}} a_j a_m\right)$ for each $j\neq m$. Because the overall factor ${1\over \prod a_j^2}$, to have
singular behavior under the soft limit, in the expansion of determinant we only need to focus on terms with at most one factor ${\Spbb{j|m}\over
\Spaa{j|m}} a_j a_m$.

For terms without such factor, the sum is nothing, but the determinant of the matrix after crossing all terms ${\Spbb{j|m}\over \Spaa{j|m}} a_j
a_m$ in diagonal elements. It is easy to see that its determinant vanishes since the sum of each row or each column is zero.

For terms with one such factor, the sum is given by the determinant of the matrix, after deleting the $m$-th and $j$-th rows and columns from
$\Psi$, and then crossing all terms ${\Spbb{t|m}\over \Spaa{t|m}} a_t a_m$ in diagonal elements with $t\neq j,m$,
\bea {\Spbb{j|m}\over \Spaa{j|m}} a_j a_m{1\over \prod_{j=1}^m a_j^2}|\Psi(\Spbb{*|m}\to 0)|_{mj}^{mj} & = &  {\Spbb{j|m}\over \Spaa{j|m}}
{a_j\over a_m} {1\over \prod_{j=1}^{m-1} a_j^2}|\Psi(\Spbb{*|m}\to 0)|_{mj}^{mj} \nn & = &  {\Spbb{j|m}\over \Spaa{j|m}} {a_j\over
a_m}h(x,M-\{m\}, y)\eea
where at the second line we have used the definition \eref{half-soft-matrix} for $m-1$ elements $M-\{m\}$.  Summing over $j$ we have shown the
soft limit behavior \eref{half-soft-factor}.

The second property we will discuss is the recursion relation
\bea \sum_{A\subset C, B= C-A} h (q, A, r) h(q, B, r) \Spaa{q|K_A K_B|q} \Spaa{r|K_A K_B| r} = -K_C^2 h(q, C, r)~~~\label{h-recur}\eea
where the summation is over all inequivalent splitting of the set $C$ into two non-empty subsets $A, B$, i.e., $(A,B)=(B,A)$. Formula
\eref{h-recur} has been proved in \cite{Bern:1998sv} and here we will give another proof.

The third property is following identity discussed in~\cite{Dunbar:2012aj},
\bea \sum_{M} h(a,M+\{c\}, b) h(b, N+\{d\}, a)=\sum_{M} h(c, M+\{a\},d) h(d, N+\{b\}, c)~~~~\label{h-rel}\eea
where the summation over $M$ is over all subsets of $\{1,2,...,n\}-\{a,b,c,d\}$ and $N=\{1,2,...,n\}-\{a,b,c,d\}-M$. This identity has not been
proved in literature and we will present a proof in this note. It is also important to notice that unlike formula \eref{h-recur}, which is true
without momentum conservation condition, identity \eref{h-rel} holds only when $\sum_{i=1}^n k_i=0$.

\subsubsection{The proof of recursion relation}

We now prove the recursion relation~\eref{h-recur} for half-soft function $h$ inductively. Using \eref{half-soft-matrix} and $A\bigcup B=C$, we
can get rid of overall factor and using the matrix form to write it as
\bea \sum_{A\subset C, B= C-A} ||\Psi_A||~~~ ||\Psi_B|| \Spaa{q|K_A K_B|q} \Spaa{r|K_A K_B| r} = -K_C^2 ||\Psi_C||~,~~\label{h-recur-1}\eea
where we have used $\Psi_A$ to denote the matrix constructed using elements in subset $A$ according to the theorem, i.e., the formula
\eref{matrix-psi} and similarly for $\Psi_B, \Psi_C$. 

For $C=\{1,2\}$, we have only one term in the sum. Using $||\Psi_A||=1$ when there is only one element in the set $A$, the left handed side
(LHS) of \eref{h-recur-1} is given by $1\times 1\times \Spaa{q|1|2|q}\Spaa{r|1|2|r}= \Spbb{1|2}^2 a_1 a_2$, which is indeed the same as the
right handed side (RHS) $-s_{12} {\Spbb{1|2}\over \Spaa{1|2}} a_1 a_2= \Spbb{1|2}^2 a_1 a_2$.

For general case with $n$ elements in $C$, we do the deformation
\bea \ket{1}\to \ket{1}-z\ket{q} ~,~~\label{h-prop1-def}\eea
which is possible since there is no momentum conservation. Under the deformation both sides of \eref{h-recur-1} will be rational function of $z$
and we consider following contour integration for $f(z)$ to be either LHS or RHS,
\bea \oint {dz\over z} {f(z)\over \Spaa{1-z q|r}^2}~.~~\label{h-first-contour} \eea
Unlike the familiar BCFW method, here we have inserted the factor $\Spaa{1-z q|r}^{-2}$, which ensures the function $f(z)/ \Spaa{1-z q|r}^2$ has
vanishing boundary contribution. Thus if we can show at all finite poles residues of $f(z)/ \Spaa{1-z q|r}^2$ are same, we prove the relation
$f_L(z=0)=f_R(z=0)$.

Now we calculate residues of various poles. First we consider single pole coming from factor $\Spaa{i|1-zq}=0$ for $i=2,...,n$.  It is easy to
see that residue at the RHS of \eref{h-recur-1} is given by\footnote{In fact, there are also factor $\Spbb{i|1} a_i \Spaa{1|q}\Spaa{1-z_i q|r}$
coming from the numerator of combination ${\Spbb{i|1}\over \Spaa{i|1-z q}} a_i a_1(z)$ as well as factor $z_i\Spaa{1-z_i q|r}^2$ coming from
denominator of \eref{h-first-contour}. But since they are the same for both sides of \eref{h-recur-1}, we drop them for simplicity. }
\bea R_i & = & -( K_C^2-z_i \Spab{q|K_C|1}) |\Psi_C(z_i)|_{1i}^{1i}~ \eea
with $z_i={\Spaa{i|1}\over \Spaa{i|q}}$. To understand $|\Psi_C(z_i)|_{1i}^{1i}$, let us notice that for $j\neq i$, two terms of its $j$-th
diagonal element can be simplified as
\bea  {\Spbb{j|1}\over \Spaa{j|1-z_i q}} a_j \Spaa{1|q}\Spaa{1-z_i q|r} + {\Spbb{j|i}\over \Spaa{j|i}} a_j a_i = {\Spbb{j|\WH i}\over
\Spaa{j|i}} a_j a_i,~~~~~~~~\bket{\WH i}= \bket{i}+\bket{1} {\Spaa{1|q}\over \Spaa{i|q}}~.~~\label{jj-comb} \eea
Thus we have
\bea |\Psi_C(z_i)|_{1i}^{1i}= |\Psi_{C-\{1\}}(\WH i)|_{i}^{i}= || \Psi_{C-\{1\}}(\WH i)||~,\eea
where $\Psi_{C-\{1\}}$ is the matrix constructed by  $n{-}1$ particles according to the graph-theoretical rule (see \eref{matrix-psi}) with the
anti-holomorphic spinor of $k_i$ shifted. Using the shifted momentum $\WH k_i$ we can see
\bean ( K_C^2-z_i \Spab{q|K_C|1})= {\Spaa{i|(K_C-k_1)K_C|q}\over \Spaa{i|q}}= (\WH K_{C-\{1\}})^2~,\eean
so finally we have residue of the RHS
\bea R_i = - (\WH K_{C-\{1\}})^2~|| \Psi_{C-\{1\}}(\WH i)||~.~~\label{h-first-R-residue}\eea

For the LHS, since we have taken the convention that $1\in A$,  to have the pole at $z_i$, we must have $i\in A$ and the residue is given by
\bea L_i&= & \sum_{1,i\in A}|\Psi_A(z_i)|_{1i}^{1i} ~~~ ||\Psi_B|| \Spaa{q|K_A K_B|q} (\Spaa{r|K_A K_B| r}-z_i\Spaa{r|q}\Spba{1|K_B|r})\nn
& = & \sum_{1,i\in A}||\Psi_{A-\{1\}}(\WH i)|| ~~~ ||\Psi_B|| \Spaa{q|( K_{A-\{1,i\}}+\WH k_i) K_B|q}\Spaa{r|( K_{A-\{1,i\}}+\WH k_i)
K_B|r}~,~~\label{h-first-L-residue}\eea
where relation $k_i+k_1- \ket{i} \bket{\WH i}= \ket{q}\bket{1} z_i$ as well as \eref{jj-comb} have been used. Comparing \eref{h-first-R-residue}
and \eref{h-first-L-residue}, and using the induction, we see immediately that residues of pole $z_i$ are same for both sides of
\eref{h-recur-1}.

Having discussed poles coming from $\Spaa{i|1-z q}$, there is only one pole left, i.e., the one coming from the inserted factor
$\Spaa{r|1-zq}^{-2}$. Naively the pole at $z_r= {\Spaa{1|r}\over \Spaa{q|r}}$ is a double pole, however, with the factor
$a_1(z)=\Spaa{q|1}\Spaa{r|1-z q}$ in combinations like ${\Spbb{i|1}\over \Spaa{i|1-zq}} a_1(z) a_i$, we must be careful in our discussion. Let
us expand the determinant $||\Psi_C||=|\Psi_C(z))|_1^1$, it is easy to see that only following two kinds of terms can have nonzero residues at
this pole:
\begin{itemize}

\item Terms without any $z$-dependence:
their summation is the determinant of a matrix obtained by removing all terms like ${\Spbb{i|1}\over \Spaa{i|1-z q}} a_1(z) a_i$ from the
diagonal of $(\Psi_C(z))_1^1$, which is just the matrix $\Psi_{C-\{1\}}$. The determinant vanishes since the sum of each row or each column is
zero.

\item Terms with only one factor
like ${\Spbb{i|1}\over \Spaa{i|1-z q}} a_1(z) a_i$ from just one diagonal element: the sum is \bea \sum_{i=2}^n  {\Spbb{i|1}\over \Spaa{i|1-z
q}} a_1(z) a_i |\W \Psi_C|_{1i}^{1i}~,\eea where $(\W \Psi_C)_{1i}^{1i}$ is obtained from matrix $\Psi_C$ by removing the first and $i$-th rows
and columns, and then removing all terms like ${\Spbb{i|1}\over \Spaa{i|1-z q}} a_1 a_i$ from the diagonal, which is the matrix $\Psi_{C-\{1\}}$
constructed from elements $\{2,3,...,n\}$. Using $|\W \Psi_C|_{1i}^{1i}=|\Psi_{C-\{1\}}|_i^i= ||\Psi_{C-\{1\}}||$ we can write it as \bea
||\Psi_{C-\{1\}}||\sum_{i=2}^n  {\Spbb{i|1}\over \Spaa{i|1-z q}} a_1 a_i~.\eea

\end{itemize}

Having understood this, we can find residue at $z_r$ for RHS as\footnote{Again we have dropped some identical factors at both sides when we
evaluate the residue.}
\bea  R_r&= & -(K_C^2-z_r\Spab{q|K_C|1})||\Psi_{C-\{1\}}||\sum_{i=2}^n {\Spbb{i|1}\over \Spaa{i|1}-z_r\Spaa{i|q}} \Spaa{1|q} a_i\nn
& = & \left( K^2_{C-\{1\}} - {\Spaa{q|1}\Spab{r|K_C|1}\over \Spaa{q|r}}\right) ||\Psi_{C-\{1\}}|| \Spaa{r|q} \Spab{q|K_C|1}\nn
& = & K^2_{C-\{1\}} ||\Psi_{C-\{1\}}|| \Spaa{r|q} \Spab{q|K_C|1}+||\Psi_{C-\{1\}}|| \Spaa{q|1}\Spab{r|K_C|1} \Spab{q|K_C|1}~.~~\label{pole-1r-R}
\eea

Now we do similar analysis for the LHS where $z$-dependence coming from $||\Psi_A(z)||$ as well as $\Spaa{r|K_A(z)|K_B|r}$. There are two cases
we need to consider separately. The first is that $A=\{1\}$. In this case, 
the residue is given by \bea  L_{r,2}=||\Psi_{C-\{1\}}|| \Spaa{q|1}\Spba{1|K_C|q} \Spba{1|K_C|r}~,\eea
which is exactly the second term at the RHS of \eref{pole-1r-R}.

For the case where $A$ has more than one element, using the same analysis as given for the set $C$, the residue is given by
\bea L_{r,1}&= & -\sum_{A-\{1\}} ||\Psi_{A-\{1\}}||~~||\Psi_{B}|| \Spaa{r|q}\Spab{q|K_A|1} \Spaa{q|K_A K_B|q}\Spaa{r|K_{A-\{1\}} K_B|r}\nn
& = & -\sum_{A-\{1\}} ||\Psi_{A-\{1\}}||~~||\Psi_{B}|| \Spaa{r|q}\Spab{q|K_A|1} \Spaa{q|K_{A-\{1\}} K_B|q}\Spaa{r|K_{A-\{1\}} K_B|r}\nn
& & - \sum_{A-\{1\}} ||\Psi_{A-\{1\}}||~~||\Psi_{B}|| \Spaa{r|q}\Spab{q|K_{A-\{1\}}|1} \Spaa{q|1}\Spba{1| K_B|q}\Spaa{r|K_{A-\{1\}} K_B|r}~.\eea
To go further we need to consider following important point. Assuming we have split remaining elements $C-\{1\}$ into two groups $\W A, \W B$,
for recursion relation of $n{-}1$ elements, $(\W A, \W B)=(\W B, \W A)$, i.e., there is only one term in the summation. However, for recursion
relation of $n$ elements, $(1\bigcup\W A, \W B)\neq (1\bigcup\W B, \W A)$, i.e., there are two terms in the summation. Using this observation,
we sum up these two terms. The first line of $L_{r,1}$  will give
\bea & & -\sum_{\W A} ||\Psi_{\W A}||~~||\Psi_{\W B}|| \Spaa{r|q}\Spab{q|K_{\W A}+K_{\W B}|1} \Spaa{q|K_{\W A} K_{\W B}|q}\Spaa{r|K_{\W A} K_{\W
B}|r}\nn & = &  K^2_{C-\{1\}} ||\Psi_{C-\{1\}}|| \Spaa{r|q} \Spab{q|K_C|1}~,\eea
where at the second line we have used the induction for $n{-}1$ elements. It is the first term at the RHS of \eref{pole-1r-R}. The second line
of $L_{r,1}$ will sum to zero because  $\Spaa{r|K_{\W A} K_{\W B}|r}+ \Spaa{r|K_{\W B} K_{\W A}|r}=0$. Having shown that all residues of finite
poles are same at two sides by induction, we have proved the recursion relation \eref{h-recur-1}.

\subsubsection{The proof of square relation}

Now we give a proof of square identity \eref{h-rel}, which is conjectured in \cite{Dunbar:2012aj} and numerically checked up to twelve points.
This relation is crucial to write the rational part of MHV one-loop amplitude of ${\cal N}=4$ supergravity theory into diagrammatic expression.

The proof of \eref{h-rel} is, in fact, quite simple if we use the matrix form and the theorem. Let us start with the $n\times n$ matrix $\Phi$
\eref{Phi-H} of the Hodges' form with arbitrary auxiliary spinors $x,y$ and  calculate the minor obtained by removing, for example, the first
four rows and four columns
\bea |\Phi|_{1234}^{1234} = {1\over \prod_{i=5}^n (\Spaa{i|x}\Spaa{i|y})^2} |\Psi|_{1234}^{1234}~~~~\label{Phi-Psi-four}\eea
where $\Psi$ is the matrix given by \eref{Psi-H}. Momentum conservation makes the definition of matrix $\Phi$ independent of the choice of
auxiliary spinors $x,y$, so is the determinant $|\Phi|_{1234}^{1234}$ at the LHS of \eref{Phi-Psi-four}.

Now we can evaluate the LHS of \eref{Phi-Psi-four} by computing the RHS by two different approaches. We can take $x=1, y=2$, so the first two
rows and first two columns of $\Psi$ are zero. The reduced $n{-}2\times n{-}2$ matrix $(\Psi)_{12}^{12}$ is exactly the form given by the
theorem, and the minor $|\Psi_{12}^{12}|_{34}^{34}$ is given by weighted forests with $3,4$ as roots. Given the set $M$ attached to root $3$ and
set $N$ attached to root $4$, all possible ways of $M$ attaching to $3$ are given by the trees calculated by $h(1,\{ M, 3\}, 2)$ up to an
overall factor, and similarly trees with $N$ attached to root $4$ are calculated by $h(1,\{ N, 4\}, 2)$. Summing over all possible $M,N$ gives
$\sum_M h(1,\{ M, 3\}, 2) h(1,\{ N, 4\}, 2)$ in \eref{h-rel}, as the forests represented by $|\Psi_{12}^{12}|_{34}^{34}$.

Alternatively, we can take $x=3, y=4$ and get the matrix $(\Psi)_{34}^{34}$. Then using the theorem, $|\Psi_{34}^{34}|_{12}^{12}$ calculates the
forests with $1,2$ as roots, which is exactly $\sum_M h(3,\{M,1\},4) h(3,\{N, 2\},4)$ in \eref{h-rel}. Since both approaches calculate the same
object, as given by the LHS of \eref{Phi-Psi-four}, we have proved the identity.

\subsection{The soft-lifting function}

Having understood the half-soft function $h$ both from the point of view of graphs and determinants, we consider the soft-lifting function
defined in~\cite{Dunbar:2012aj}. With a little bit of algebra, it is easy to see that up to a factor, the soft-lifting function $S[P^s;
Q^p]_{m_1, m_2}$ with $s$ elements in the set $P$ and $p$ elements in the set $Q$ is, in fact, weighted forests of the set $P\bigcup Q$ with
roots given by all elements in the set $P$ (where $m_1, m_2$ are auxiliary spinor for the factor $\Spaa{i|m_1}\Spaa{i|m_2}=a_i$). More
precisely,
\bea S[P^s; Q^p] & = & {1\over \prod_{t\in  Q} a_t^2} \sum_{\textrm{forest}(p_1,...,p_s)} \prod_{\textrm{edges}~(rs)} {\Spbb{r|s}\over
\Spaa{r|s}} a_r a_s \nn
& = &  {1\over\prod_{t\in Q} a_{t}^2}|\Psi|^{p_1 ... p_s}_{p_1...p_s}=|\Phi|^{p_1 ... p_s}_{p_1...p_s} ~~~~\label{lift-matrix}\eea
where we have used the notation $\textrm{forest}(p_1,...,p_s)$ for forest of set $P\bigcup Q$ with $p_1,...p_s$ as roots. At the second line we
have used both matrix forms $\Psi,\Phi$ to write down the expression using the determinant.

Using the matrix form \eref{lift-matrix}, it is easy to see the soft behavior of soft-lifting function as
\bea S[P^s; Q^p] & \to & -{\cal S}_{q^+}(m_1, P^s\bigcup Q^p-\{q\}, m_2) S[ P^s; Q^p-\{ q\}]~~~\label{soft-lifting-soft}\eea
and there is no soft-singularity if $q_1\not\in Q^p$. The reason is that under the soft limit, $\Spbb{r|s}/\Spaa{r|s}$ is a smooth function,
thus all singular behaviors come from the overall factor ${1\over\prod_{t\in Q} a_{t}^2}$. Using the same arguments for soft limit of half-soft
function \eref{h-soft-limit}, we can show \eref{soft-lifting-soft}.

\section{The rational part of one-loop MHV ${\cal N}=4$ supergravity
amplitudes}\label{one-loop} 

\begin{figure}\centering
\includegraphics[height=5cm]{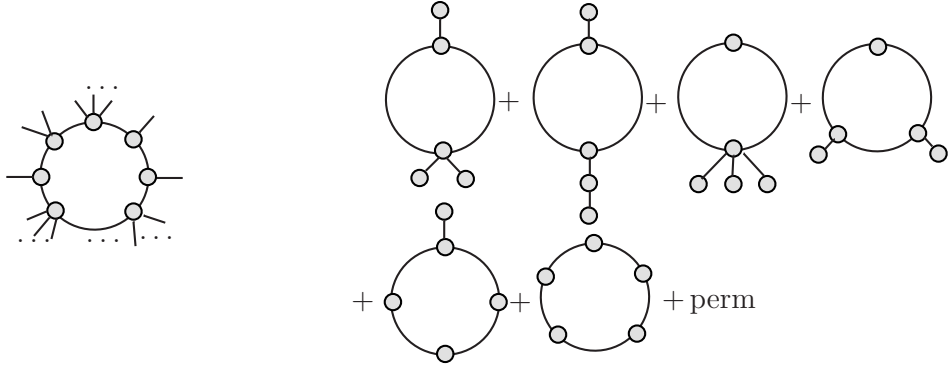}
\caption{Diagrammatic formula for the rational part of one-loop MHV amplitudes. A general connected, one-loop diagram (left) is a loop with $r$
nodes, attached with forests, which have $r$ trees containing the $r$ nodes as their roots. The 7-pt case (right): we pick two reference labels,
and sum over all possible assignments of the remaining $n{-}2$ ones to the vertices.} \label{fig:one_loop}
\end{figure}

Recently, the rational part of one-loop MHV ${\cal N}=4$ supergravity amplitudes has been calculated and conjectured in
\cite{Dunbar:2010fy,Dunbar:2011dw,Dunbar:2012aj} based on the identity \eref{h-rel} proved in this note. Using the identity \eref{h-rel} they
obtained an expression for MHV one-loop rational function,
\bea {\cal R}_n^{\textrm{MHV}} & = & {(-)^n \Spaa{1|2}^4\over \prod_{i=3}^n (\Spaa{i|1}\Spaa{i|2})^2} \sum_{\textrm{1-loop}}
\prod_{\textrm{edge} (ab)} {\Spbb{a|b}\over \Spaa{a|b}}\Spaa{b|1}\Spaa{b|2}\Spaa{a|1}\Spaa{a|2}~~~~\label{R-graph-def}\eea
where the sum is over all distinct, connected, one-loop ``link diagrams'' with $(n-2)$-nodes, see Fig.~\ref{fig:one_loop}. Along the loop, there
can be $r=2,...,n-2$ nodes. For given $r$ nodes along the loop, there are ${(r-1)!\over 2}$ different ordering, which means the diagram is not
directed (clockwise ordering will be identified with anti-clockwise ordering). Also it is important to notice that while for general $r\geq 3$
the weight of graphes is one, when $r=2$, the weight is ${1\over 2}$.

With the understanding of Hodges' determinant using matrix-tree theorem, it is natural to ask if we can derive an determinant for this
diagrammatic expansion~\eref{R-graph-def}. In this section, we will propose one expression (partially) realize this idea.

\subsection{A proposal for the matrix}

Here we propose our matrices, whose determinant will contain the one-loop result presented in \eref{R-graph-def},
%
\bea \WH \Phi=\left( \begin{array}{ccccc} -\sum_{j\neq 1} \phi_1^j {a_j\over a_1} &  (1+q e_{12})\phi_1^2  & \ldots   &  (1+q
e_{1(n-1)})\phi_1^{n-1}
 &  (1+q e_{1n})\phi_1^n \\
 (1+q e_{21})\phi_2^1 & -\sum_{j\neq 2} \phi_2^j {a_j\over a_2} &   \ldots  &
 (1+q e_{2(n-1)})\phi_2^{n-1} &
(1+q e_{2n})\phi_2^n  \\
\\ \vdots &\vdots & \ddots  & \vdots &\vdots \\
(1+q e_{(n-1)1})\phi_{n-1}^1 & (1+q e_{(n-1)2})\phi_{n-1}^2 & \ldots & -\sum_{j\neq n-1} \phi_{n-1}^j {a_j\over a_{n-1}}
 & (1+q e_{(n-1)n})\phi_{n-1}^{n}\\
(1+q e_{n1})\phi_{n}^1 & (1+q e_{n2})\phi_{n}^2 &  \ldots & (1+q e_{n(n-1)})\phi_n^{n-1} & -\sum_{j\neq n} \phi_{n-1}^j {a_j\over a_{n}}
\end{array} \right)~~~\label{Guess-Phi}\eea
and a related, more symmetric matrix,
\bea \WH \Psi= A\cdot \WH \Phi\cdot A~~~~~\label{Guess-Psi}\eea

Compared with \eref{Phi-H} we have almost same form except addressing factor $(1+q e_{ij})$ for off-diagonal elements. The consequence of this
factor is following:
\begin{itemize}

\item The matrix $\WH \Phi$ is still independent of the choice
of auxiliary spinors $x,y$, but it is not degenerate anymore, i.e., its rank is $n$.

\item The role of $q$ is to count the number of vertices along
loops. In other words, coefficient of $q^m$ are related to diagrams with $m$ vertices along loops (it could be one or multiple loops).

\item The role of $e_{ij}$ is to give information about loops in the
diagrams, so we need to impose following algebraic rule:
\bea e_{ij} e_{jk}= e_{ik},~~~~~e_{ij} e_{kl}=e_{kl} e_{ij} (i\neq j\neq k\neq l),~~~~e_{ii}=N_c~~~\label{e-rule}\eea

\end{itemize}

Using above two matrices we can calculate following expression
\bea {\cal W}_{n;P} & = & |\WH \Phi|_{p_1 p_2...p_r}^{p_1 p_2...p_r}={1\over \prod_{i=1, i\not\in P}^n a_i^2} |\WH \Psi|_{p_1 p_2...p_r}^{p_1
p_2...p_r} ~~~~\label{Guess-obj-more-1} \eea
where the set $P$ tells us which rows and columns have been removed. Since matrix $\WH \Phi$ is $x,y$-independent, any object given in
\eref{Guess-obj-more-1} is also $x,y$-independent. Thus we can choose $x,y$ to simplify the calculation.

Now we give a prescription to read out one-loop rational contributions from ${\cal W}_{n;\{1,2\}}$ assuming only $1,2$ have negative helicities
and all others, positive helicities.
\begin{itemize}

\item (1) Taking the determinant, i.e., calculating ${\cal W}_{n;\{1,2\}}$.

\item (2) Applying \eref{e-rule} iteratively until we can not
simplify further.

\item (3) Taking all $e_{ij}\to 0$. This will get rid of all
unwanted terms.

\item (4) Now we have a polynomial of the form $\sum_{m=2}^{n-2} \sum_{t=1}^{[(n-2)/2]} q^m N_c^t f_{m,t}$, where the power $t$ counts the number of loops in a given diagram, and
the power $m$ counts the number of nodes located along the $t$ loops.

\item (5)
For the one-loop result, the answer is the
sum $\sum_{m=2}^{n-2} f_{m,1}$ multiplying by $(-1)^n \Spaa{1|2}^4/2$. In other words, we just need the $t=1$ case and set $q=1$.

\end{itemize}
%
\subsection{A few terms in the expansion}

Having the above prescription, let us check a few terms in the expansion of determinant. To compare with results given in
\cite{Dunbar:2010fy,Dunbar:2011dw,Dunbar:2012aj}, we will choose $x=1, y=2$ and expand the determinant according to the power of $q$.
\begin{itemize}

\item {\bf Terms with $q^0$:} In this case, we set all $q=0$ and the
matrix reduces to old matrix $\Psi$. Since its rank is $n{-}3$, the determinant vanishes.

\item {\bf Terms with $q^1$:} In this case, we should take one
term with $q$ from, for example, the off-diagonal element with $i\neq j$ and then set all other $q=0$. However, since we need to set the
$e_{ij}\to 0$, the result is zero too.

\item {\bf Terms with $q^2$:} It is easy to see that all terms will be following form
\bea \sum_{(ij), (kl)}(-)^{i+j+k+l}(q e_{ij} \phi_i^j a_i a_j) (q e_{ kl}\phi_k^l a_k a_l)|\Psi|_{ik}^{jl}~~~\label{Dunbar-q2-part-1} \eea
where we sum over two pairs of indices with condition that $i\neq j, k$ and $l\neq k,j$ and $|\Psi|_{ik}^{jl}$ is the minor after removing the
$i$-th, $k$-th rows and columns. Depending on the choice of pairs we have several cases. For the case with $i\neq l$ and $j\neq k$, or the case
with $i= l$ or $j= k$, no $e_{ij}$ is left after using \eref{e-rule}, so there is no such contributions.

The case $i= l$ and $j= k$ gives following result,
    \bea \sum_{ (ij), i\neq j} N_c (  \phi_i^j a_i a_j)^2|\Psi|_{ij}^{ij}~,~~\label{Dunbar-q2-part-3}\eea
after applying our prescriptions. The graphic picture of minor $|\Psi|_{ij}^{ij}$ will be forest
    with two roots $i,j$ and  then we connect $i,j$  by two edges. They are exactly graphes of one-loop rational terms with nodes $i,j$ along the loop.

\item {\bf Terms with $q^3$:} The result is
\bea \sum_{ (i_1,j_1), (i_2, j_2), (i_3,j_3)}(-)^{\sum_{t=1}^3 (i_t+j_t)} q^3 e_{i_1 j_1} e_{i_2 j_2} e_{i_3 j_3}\psi_{i_1}^{j_1}
\psi_{i_2}^{j_2} \psi_{i_3}^{j_3}|\Psi|_{i_1 i_2 i_3}^{j_1 j_2 j_3}~~~\label{Dunbar-q3-part} \eea
where $i_1\neq i_2\neq i_3$ and $j_1\neq j_2\neq j_3$ and $i_t\neq j_t, t=1,2,3$.

Again we have various cases, with 0,1,2 or 3 common indices between the sets $(i_1, i_2, i_3)$ and $(j_1, j_2, j_3)$. As it is clear from our
prescription, when we set $e_{ij}\to 0$, only the last case (when the two sets are related by permutations) survives, which has $e_{ij}$ left
after applying \eref{e-rule}. The contribution is,
\bea 2\sum_{3\leq i_1<i_2<i_3\leq n-2} q^3 N_c \psi_{i_1}^{i_2} \psi_{i_2}^{i_3} \psi_{i_3}^{i_1}|\Psi|_{i_1 i_2 i_3}^{i_1 i_2
i_3}~~~\label{Dunbar-q3-part-1}\eea
It is important to notice the factor $2$ in \eref{Dunbar-q3-part-1}. Same factor $2$ will appear in all $q^{m\geq 3}$ case. The pattern that,
compared to $q^2$ term, there is an additional factor of $2$ for $q^{m\geq 3}$ terms, agrees exactly with that in~\cite{Dunbar:2012aj}.

\item {\bf Terms with general $q^{m\geq 3}$:} General terms are given by
\bea \sum_{I\subset \{3,..,n\}} |L(I)|_{N_c}~~
|\Psi|_{I}^I~~~\label{Loop-rule-gen-form}\eea
where we sum over all distinct subsets of $\{3,...,m\}$, and the matrix $L$ is defined, e.g. for $m=4$,

\bea L(i_1,i_2,i_3,i_4)\equiv \left( \begin{array}{cccc} 0 & q e_{i_1 j_2} \psi_{i_1}^{j_2} & q e_{i_1 j_3} \psi_{i_1}^{j_3} & q e_{i_1 j_4}
\psi_{i_1}^{j_4} \\ q e_{i_2 j_1} \psi_{i_2}^{j_1} & 0 & q e_{i_2 j_3} \psi_{i_2}^{j_3} & q e_{i_2 j_4} \psi_{i_2}^{j_4} \\ q e_{i_3 j_1}
\psi_{i_3}^{j_1} & q e_{i_3 j_2} \psi_{i_3}^{j_2} & 0 & q e_{i_3 j_4} \psi_{i_3}^{j_4}
\\ q e_{i_4 j_1}
\psi_{i_4}^{j_1} & q e_{i_4 j_2} \psi_{i_4}^{j_2} & q e_{i_4 j_3} \psi_{i_4}^{j_3} & 0
\end{array}\right)~.~~~\label{loop-q4-part-loop}\eea
The sub-index $N_c$ means we keep only the term with power $N_c^1$.

Given the set $I$, $|L(I)|_{N_c}$ describes how nodes are
distributed along the loop while $|\Psi|_{I}^I$ describes how forest
are constructed with roots on the set $I$. Also, it is easy to see
why there are weights $1$ and $1/2$ for $q^{a\geq 3}$ and $q^2$
respectively. With three or more nodes along the loop, we can
distinguish the clockwise or anti-clockwise ordering while with two
nodes, there is only one ordering.

\end{itemize}
%

\section{Final remarks}

Based on the matrix-tree theorem, in this note we explore the connection between graphs and determinants which appear naturally in gravity
amplitudes. For MHV tree amplitudes, it is straightforward to identify Hodges' determinant with NSVW's tree diagrams, and we have learnt about
its most general graphic structures. Given that non-MHV amplitudes can also be expressed in terms of
determinant~\cite{Cachazo:2012da}~\cite{Cachazo:2012kg}, we expect similar diagrammatic formulations for all tree amplitudes. For example, with
two determinants used in \cite{Cachazo:2012kg}, the formula can be expanded into product of two spanning trees.

We have studied some universal functions for constructing gravity amplitudes in graph/determinant formulations. These includes half-soft and
soft-lifting functions, and we prove non-trivial identities using the formulations. We have also proposed a matrix to calculate the one-loop
rational terms in $\mathcal{N}=4$ supergravity. The proposal is not the most satisfying in the sense that some prescriptions like \eref{e-rule}
are needed. The expansion of the determinant has too many terms, which contains not only one-loop, but also higher-loop structures. Given the
interesting connection between MHV amplitudes at tree and one-loop level (for rational part), it is natural to ask if these higher loop
structures are related to certain higher loop rational contributions.

We find both formulations useful for understanding gravity amplitudes, and deserve further studies. With the one-loop rational part as an
example, it would be fascinating to explore similar structures for supergravity loop amplitudes in general. Eventually one would like to
understand the physical interpretation of these formulations, which might be provided by a twistor-string/Grassmannian-like dual formulation of
the S-matrix in supergravity theories.


\subsection*{Acknowledgements}

We would like to thank Freddy Cachazo,  Andrew Hodges, David Skinner for discussions, and Yu-xiang Gu for checking some calculations. We thank
Perimeter Institute for its hospitality, where the project was started, and B.~Feng also thanks Niels Bohr International Academy and Discovery
Center, where the project is finished. B.~Feng is supported by fund from Qiu-Shi, as well as Chinese NSF funding under contract No.11031005,
No.11135006, No. 11125523. S. He's visit to Perimeter Institute is partially supported by the EC ``UNIFY'' grant.



\begin{thebibliography}{References}

\bibitem{ampreview}
 For recent reviews, see: 
 R.~Roiban, M.~Spradlin and A.~Volovich (Ed),
 J.\ Phys.\ A: Math. Theor. 44 450301 (2011).

\bibitem{Britto:2004ap}
  R.~Britto, F.~Cachazo and B.~Feng,
  Nucl.\ Phys.\  B {\bf 715}, 499 (2005)
  [arXiv:hep-th/0412308].

\bibitem{Britto:2005fq}
  R.~Britto, F.~Cachazo, B.~Feng and E.~Witten,
  Phys.\ Rev.\ Lett.\  {\bf 94}, 181602 (2005)
  [arXiv:hep-th/0501052].

\bibitem{Bern:1994zx}
  Z.~Bern, L.~J.~Dixon, D.~C.~Dunbar and D.~A.~Kosower,
  Nucl.\ Phys.\ B {\bf 425}, 217 (1994)
  [hep-ph/9403226].

\bibitem{Bern:1994cg}
  Z.~Bern, L.~J.~Dixon, D.~C.~Dunbar and D.~A.~Kosower,
  Nucl.\ Phys.\ B {\bf 435}, 59 (1995)
  [hep-ph/9409265].

\bibitem{Grassmannian}
  N.~Arkani-Hamed, F.~Cachazo, C.~Cheung, J.~Kaplan,
  JHEP {\bf 1003}, 020 (2010).
  [arXiv:0907.5418 [hep-th]].

\bibitem{intreview}
 For recent reviews, see: N.~Beisert, C.~Ahn, L.~F.~Alday, Z.~Bajnok, J.~M.~Drummond, L.~Freyhult, N.~Gromov, R.~A.~Janik {\it et al.},
   [arXiv:1012.3982 [hep-th]].

\bibitem{all-loop}
 See e.g. N.~Arkani-Hamed, J.~L.~Bourjaily, F.~Cachazo, S.~Caron-Huot and J.~Trnka,
  JHEP {\bf 1101}, 041 (2011)
  [arXiv:1008.2958 [hep-th]].
  S.~Caron-Huot and S.~He,
  arXiv:1112.1060 [hep-th].

\bibitem{Kawai:1985xq}
  H.~Kawai, D.~C.~Lewellen and S.~H.~H.~Tye,
  Nucl.\ Phys.\ B {\bf 269}, 1 (1986).

\bibitem{Bern:2008qj}
  Z.~Bern, J.~J.~M.~Carrasco and H.~Johansson,
  Phys.\ Rev.\  D {\bf 78}, 085011 (2008)
  [arXiv:0805.3993 [hep-ph]].

\bibitem{Bern:2010yg}
  Z.~Bern, T.~Dennen, Y.~-t.~Huang and M.~Kiermaier,
  Phys.\ Rev.\ D {\bf 82}, 065003 (2010)
  [arXiv:1004.0693 [hep-th]].

\bibitem{BCJstring}
  N.~E.~J.~Bjerrum-Bohr, P.~H.~Damgaard and P.~Vanhove,
  Phys.\ Rev.\ Lett.\  {\bf 103}, 161602 (2009)
  [arXiv:0907.1425 [hep-th]].
  S.~Stieberger,
  arXiv:0907.2211 [hep-th].
  H.~Tye and Y.~Zhang,
  arXiv:1003.1732 [hep-th].

\bibitem{BCJfieldtheory}
 B.~Feng, R.~Huang and Y.~Jia,
  arXiv:1004.3417 [hep-th].

\bibitem{KLTfieldtheory}
  N.~E.~J.~Bjerrum-Bohr, P.~H.~Damgaard, B.~Feng and T.~Sondergaard,
  arXiv:1005.4367 [hep-th].
  N.~E.~J.~Bjerrum-Bohr, P.~H.~Damgaard, B.~Feng and T.~Sondergaard,
  arXiv:1006.3214 [hep-th].
  B.~Feng and S.~He,
  arXiv:1007.0055 [hep-th].
  N.~E.~J.~Bjerrum-Bohr, P.~H.~Damgaard, B.~Feng and T.~Sondergaard,
  arXiv:1007.3111 [hep-th].

\bibitem{Bargheer:2012gv}
  T.~Bargheer, S.~He and T.~McLoughlin,
  arXiv:1203.0562 [hep-th].

\bibitem{Bern:2010ue}
  Z.~Bern, J.~J.~M.~Carrasco and H.~Johansson,
  Phys.\ Rev.\ Lett.\  {\bf 105}, 061602 (2010)
  [arXiv:1004.0476 [hep-th]].

\bibitem{SUGRAUV}
  Z.~Bern, J.~J.~Carrasco, L.~J.~Dixon, H.~Johansson, D.~A.~Kosower and R.~Roiban,
  Phys.\ Rev.\ Lett.\  {\bf 98}, 161303 (2007)
  [hep-th/0702112].
  Z.~Bern, J.~J.~M.~Carrasco, L.~J.~Dixon, H.~Johansson and R.~Roiban,
  Phys.\ Rev.\ D {\bf 78}, 105019 (2008)
  [arXiv:0808.4112 [hep-th]].
  Z.~Bern, J.~J.~Carrasco, L.~J.~Dixon, H.~Johansson and R.~Roiban,
  Phys.\ Rev.\ Lett.\  {\bf 103}, 081301 (2009)
  [arXiv:0905.2326 [hep-th]].
  Z.~Bern, J.~J.~M.~Carrasco, L.~J.~Dixon, H.~Johansson and R.~Roiban,
  Phys.\ Rev.\ D {\bf 82}, 125040 (2010)
  [arXiv:1008.3327 [hep-th]].

\bibitem{ArkaniHamed:2008gz}
  N.~Arkani-Hamed, F.~Cachazo and J.~Kaplan,
  JHEP {\bf 1009}, 016 (2010)
  [arXiv:0808.1446 [hep-th]].

\bibitem{Berends:1988zp}
  F.~A.~Berends, W.~T.~Giele and H.~Kuijf,
  Phys.\ Lett.\  B {\bf 211}, 91 (1988).

\bibitem{Nguyen:2009jk}
  D.~Nguyen, M.~Spradlin, A.~Volovich and C.~Wen,
  JHEP {\bf 1007}, 045 (2010)  [arXiv:0907.2276 [hep-th]].  

\bibitem{Hodges:2012ym}
  A.~Hodges,
   arXiv:1204.1930 [hep-th].  

\bibitem{Cachazo:2012da}
  F.~Cachazo and Y.~Geyer,
  arXiv:1206.6511 [hep-th].

\bibitem{Cachazo:2012kg}
  F.~Cachazo and D.~Skinner,
  arXiv:1207.0741 [hep-th].

\bibitem{Dunbar:2010fy}
  D.~C.~Dunbar, J.~H.~Ettle and W.~B.~Perkins,
  Phys.\ Rev.\ D {\bf 83}, 065015 (2011)
  [arXiv:1011.5378 [hep-th]].

\bibitem{Dunbar:2011dw}
  D.~C.~Dunbar, J.~H.~Ettle and W.~B.~Perkins,
  Phys.\ Rev.\ Lett.\  {\bf 108}, 061603 (2012)
  [arXiv:1111.1153 [hep-th]].

\bibitem{Dunbar:2012aj}
  D.~C.~Dunbar, J.~H.~Ettle and W.~B.~Perkin,
   arXiv:1203.0198 [hep-th].  

\bibitem{Stanleybook}
 See e.g. R.~P.~Stanley, ``Enumarative Combinatorics (Vol 2),'', Cambridge University Press, 2001.

\bibitem{Hodges:2011wm}
  A.~Hodges,
  arXiv:1108.2227 [hep-th].

\bibitem{Bern:1998sv}
  Z.~Bern, L.~J.~Dixon, M.~Perelstein and J.~S.~Rozowsky,
  Nucl.\ Phys.\  B {\bf 546}, 423 (1999)
  [arXiv:hep-th/9811140].

\end{thebibliography}
\end{document}